\documentclass[
     12pt,
     draftcls,
     journal,
     letterpaper,
     onecolumn,
]{IEEEtran}

\usepackage{epsfig,color}
\usepackage{subfigure}
\usepackage{amsmath,amssymb,bbm, dsfont}
\usepackage{float}
\usepackage{algorithmic}
\usepackage{algorithm}
\usepackage{verbatim}
\usepackage{algorithmic}
\usepackage{cite}

\newcommand{\beq}{ \begin{equation} }
\newcommand{\eeq}{ \end{equation} }
\newcommand{\beqa}{\begin{eqnarray}}%
\newcommand{\eeqa}{\end{eqnarray}}

\newcommand{\cvec}{\mbox{\boldmath $c$}}

\renewcommand{\baselinestretch}{1.7}
\begin{document}
%\sf
%\thispagestyle{empty}
\title{Power Allocation for Conventional and Buffer-Aided Link Adaptive Relaying Systems with Energy Harvesting Nodes\footnote{This paper has been presented in part at the IEEE Vehicular Technology Conference (VTC), Spring, Yokohama, Japan, May 2012.}}
\author{Imtiaz Ahmed$^\dagger$, Aissa Ikhlef$^\dagger$, Robert Schober$^\dagger$, and Ranjan K. Mallik$^{\dagger\dagger}$\\
$^\dagger$University of British Columbia, Vancouver, Canada, \\
$^{\dagger\dagger}$Indian Institute of Technology, Delhi, India\\
E-mail: \{imtiazah,\,aikhlef,\,rschober\}@ece.ubc.ca, rkmallik@ee.iitd.ernet.in}
\maketitle
\vspace{-1.8cm}
\renewcommand{\baselinestretch}{1.4}
%\small\normalsize
%\setlength{\topsep}{-2pt}
\begin{abstract}
Energy harvesting (EH) nodes can play an important role in  cooperative communication systems which do not have a continuous power supply. In this paper, we consider the optimization of conventional and buffer--aided link adaptive EH relaying systems, where an EH source communicates with the destination via an EH decode--and--forward relay. In conventional relaying, source and relay transmit signals in consecutive time slots whereas in buffer--aided link adaptive relaying, the state of the source--relay and relay--destination channels determines whether the source or the relay is selected for transmission. Our objective is to maximize the system throughput over a finite number of transmission time slots for both relaying protocols. In case of conventional relaying, we propose an offline and several online joint source and relay transmit power allocation schemes. For offline power allocation, we formulate an optimization problem  which can be solved optimally. For the online case, we propose a dynamic programming (DP) approach to compute the optimal online transmit power. To alleviate the complexity inherent to DP, we also propose several suboptimal online power allocation schemes. For buffer--aided link adaptive relaying, we show that the joint offline optimization of the source and relay transmit powers along with the link selection results in a mixed integer non--linear program which we solve optimally using the spatial branch--and--bound method. We also propose an efficient online power allocation scheme and a naive online power allocation scheme for buffer--aided link adaptive relaying. Our results show that link adaptive relaying provides performance improvement over conventional relaying at the expense of a higher computational complexity.

%Using energy harvesting nodes can be a viable solution for energy limited cooperative communication systems. In this paper, we consider the optimization of an energy harvesting (EH) relay system, where an EH source communicates with the destination via an EH decode--and--forward (DF) relay. Our objective is to maximize the system throughput over a finite number of transmission intervals. To this end, we propose an offline and several online joint source and relay transmit power allocation schemes.  For offline power allocation, we formulate a convex optimization problem which can be solved either in closed form or using standard optimization tools. For the online case, we propose a dynamic programming (DP) approach to compute the optimal online transmit power. To alleviate the complexity inherent to DP, we propose several suboptimal online power allocation schemes. Our simulation results show that the developed suboptimal schemes provide a good complexity--performance tradeoff compared to optimal online power allocation.
%%The accuracy of the derived analytical results and the effectiveness of relay subset selection
%%and power allocation are confirmed by computer simulations.
\end{abstract}

%\begin{keywords} Energy Harvesting, Decode--and--Forward, Power Allocation, Convex Optimization, Dynamic Programming.
%\end{keywords}

\newpage
\renewcommand{\baselinestretch}{1.7}
%\small\normalsize
%\setcounter{page}{1}
\pagenumbering{arabic}

%% Introduction
\section{Introduction \label{s1}}
In cooperative communication systems, a source and a number of cooperating relays expend their energy for processing and transmitting data. For some applications, connecting the source and the relays to the power grid is cumbersome or may even not be possible. Pre–-charged batteries can be a viable solution to overcome this problem. In practice, the limited storage capacity of batteries and high transmit powers may result in quick drainage of the batteries. As a result, the batteries need to be replaced/recharged periodically which can be sometimes impractical. An alternative solution is the deployment of energy harvesting (EH) nodes. EH nodes harvest energy from their surrounding environment to carry out their functions. Energy can be harvested using solar, thermoelectric, and motion effects, or through other physical phenomena \cite{paper_EH_7}. An EH node that has used up its stored energy can harvest new energy and become again available for transmission. Thus, EH nodes can be regarded as a promising option for deployment as they ensure a long system lifetime without the need for periodic battery replacements. In EH cooperative systems, the energy can be independently harvested by the EH source and/or EH relays during the course of data transmission at random times and in random amounts. For data transmission (and for other signal processing tasks), EH nodes expend the energy from their storage and only the unused energy remains in the batteries. In particular, at each time slot, the source and the relays are constrained to use at most the energy available in their storage. These constraints necessitate the design of new transmission strategies for the source and the relays to ensure optimum performance in an EH environment.

Recently, transmission strategies for and performance analyses of EH nodes in wireless communication systems have been provided in \cite{paper_EH_1,paper_EH_2,paper_EH_3,paper_EH_8,paper_EH_9,paper_EH_11}. In \cite{paper_EH_1}, a single source--destination non--cooperative link with an EH source was considered and an optimal offline along with an optimal and several sub--optimal online transmission policies were provided for allocating transmit power to the source according to the random variations of the channel and the energy storage conditions. In \cite{paper_EH_2}, a similar system model was considered, where dynamic programming (DP) was employed to allocate the source transmit power for the case when causal channel state information (CSI) was available. Several higher layer issues such as transmission time minimization and transmission packet scheduling in EH systems were considered in \cite{paper_EH_8, paper_EH_9, paper_EH_11}. The deployment of EH sensors in sensor networks has been extensively discussed in the literature \cite{paper_EH_7, paper_EH_6}. The use of EH relays in cooperative communication was introduced in \cite{paper_EH_3}, where a comprehensive performance analysis was performed for relay selection in a cooperative network employing EH relays. A deterministic EH model (assuming a priori knowledge of the energy arrival times and the amount of harvested energy) for the Gaussian relay channel was considered in \cite{paper_EH_14}, \cite{paper51}, where delay and non--delay constrained traffic were studied. The concept of energy transfer in EH relay systems was considered in \cite{paper52}, where an offline power allocation scheme was proposed.

In this paper, we consider a simple single link cooperative system where the source communicates with the destination via a decode--and--forward (DF) relay. We assume that the relay operates in half--duplex mode. In most of the existing literature on half duplex relaying \cite{paper_EH_3,paper_EH_14,paper_EH_12,paper_EH_13}, it is assumed that relays receive a packet in one time slot from the source and forward it in the next time slot to the destination. We refer to this approach as ``conventional'' relaying throughout the paper. Recently, it has been shown in \cite{paper_EH_15, paper_EH_16} that equipping relays with buffers can improve the performance of cooperative communication systems. In fact, using buffers at the relays allows storage of packets temporarily at the relay if the relay--destination channel condition is not good enough until the quality of the channel has sufficiently improved. In \cite{paper_EH_16}, a buffer--aided adaptive link selection protocol was proposed. This protocol gives relays the freedom to decide in which time slot to receive and in which time slot to transmit.

In this paper, we assume that the source and the relay are EH nodes and consider both conventional and buffer--aided link adaptive relaying. For both protocols, we propose offline and online (real--time) power allocation schemes that maximize the end--to--end system throughput over a finite number of transmission slots. The offline schemes are of interest when the amount of harvested energy and the channel signal--to--noise ratio (SNR) for all transmission slots are known a priori. However, in practice, the amount of harvested energy and the channel SNR are random in nature and cannot be predicted in advance. Therefore, in this case, online power allocation schemes have to be employed taking into account the available knowledge of channel SNR and harvested energy. Nevertheless, considering offline schemes is still important since they provide performance upper bounds for the practical online schemes. For conventional relaying, we propose an optimal online power allocation scheme which is based on a stochastic DP approach. To avoid the high complexity inherent to DP, we also propose several sub--optimal online algorithms. In case of buffer--aided link adaptive relaying, we formulate an offline optimization problem that jointly optimizes the source and the relay transmit powers along with the link selection variable. Thereby, the link selection variable indicates whether the source or the relay is selected for transmission in a given time slot. The optimization problem is shown to be a non--convex mixed integer non--linear program (MINLP). We propose to use the spatial branch--and--bound (sBB) method to solve the offline MINLP problem optimally \cite{paper46, paper47, paper49}.  We also propose a practical online power allocation scheme for the buffer--aided link adaptive relaying protocol. We note that our buffer--aided link adaptive protocol is significantly different from the delay constrained model in \cite{paper_EH_14}. The model in \cite{paper_EH_14} assumes the presence of a buffer at the relay but does not consider adaptive link selection.

The remainder of this paper is organized as follows. In Section \ref{s2}, the system model for the EH system is presented. Different power allocation schemes for conventional and buffer--aided link adaptive relaying are provided in Sections \ref{s3} and \ref{s4}, respectively. In Section \ref{s5}, the effectiveness of these power allocation schemes is evaluated based on simulations. Section \ref{s6} concludes this paper.

%However, to the best of our knowledge, the transmission power management in a cooperative communication system with EH nodes has not been considered before.

%However, we could not formulate the online power allocation problem (for buffer aided adaptive link selection protocol) by stochastic DP due to its complicated formulation. Therefore, we analytically solve an optimization problem and propose an efficient online algorithm to allocate the source and relay transmit powers.

%%%%%%%%%%%%%%%%
\section{System Model \label{s2}}
%%%%%%%%%%%%%%%%%
\begin{figure}[h]
\centering
\includegraphics[width = 3.8in]{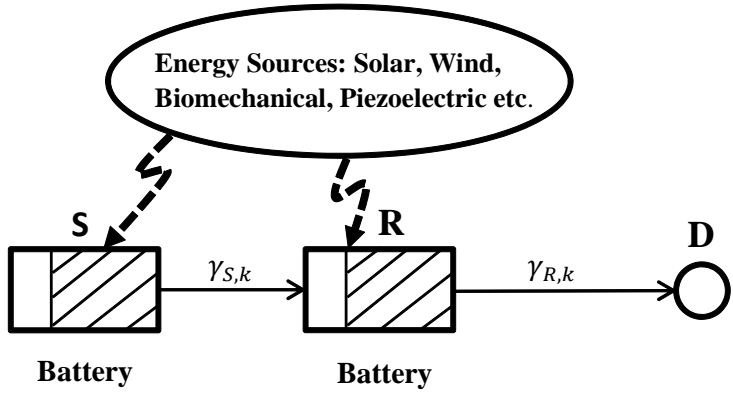} \vspace{-6mm}
 \caption{System model for a single link $S$--$R$--$D$ system where $S$ and $R$ are EH devices. $S$ and $R$ have batteries with finite storage which can store the energies harvested from the energy sources. The rectangle boxes in $S$ and $R$ represent the batteries and the hashed areas represent the amount of energy stored.} \label{figblock} \vspace{-7mm}
\end{figure}

We consider an EH relay system, where the source, $S$, communicates with the destination, $D$, via a cooperative relay, $R$, as shown in Fig.~\ref{figblock}. Both $S$ and $R$ are EH devices and their participation in signal transmission and processing depends on the harvested energy. The harvested energy can be of any form, e.g. solar, wind, or electro--mechanical energy. $S$ and $R$ are equipped with batteries, which have limited storage capacity and store the harvested energy for future use. In particular, the batteries of $S$ and $R$ can store at most $B_{S,max}$ and $B_{R,max}$ units of energy, respectively. We assume the transmission is organized in time slots of duration $T$. In the following, without loss of generality, we set $T=1$, and the system transmits for $K$ time slots. In the next two subsections, we discuss the signal model, the system throughput, and the battery dynamics for conventional and  buffer--aided link adaptive relaying.

\subsection{Conventional Relaying}
\noindent \textbf{Signal Model:} In conventional relaying, during the first time slot, $S$ transmits and $R$ receives, and during the second time slot, $R$ transmits and $D$ receives. This sequential process continues for $K$ time slots. Here, $K$ is assumed to be an even number. The received packet at $R$ in the $(2k-1)$th time slot, where $k \in \{1,2,\cdots,K/2\}$, is modelled as
\beqa y_{R,2k-1} = h_{S,2k-1}x_{2k-1} + n_{R,2k-1} \label{eq_A_1},\eeqa
where $h_{S,2k-1}$ is the fading gain of the $S$--$R$ link, and $n_{R,2k-1}$ denotes the noise sample at $R$. The transmitted packet $x_{2k-1}$ contains Gaussian--distributed symbols.
Assuming DF relaying, the detected packet, $\hat x_{2k}$, is transmitted from $R$ during time slot $2k$. Thus,
the received packet at $D$ is given by
\beqa y_{D,2k} = h_{R,2k} \hat x_{2k} + n_{D,2k} \label{eq_A_2},\eeqa
where $h_{R,2k}$ and $n_{D,2k}$ denote the fading gain of the
$R$--$D$ link and the noise sample at $D$, respectively. $h_{S,2k-1}$ and $h_{R,2k}$ can follow any fading distribution, e.g. Rayleigh, Rician, Nakagami--$q$, or Nakagami--$m$ fading. $n_{R,2k-1}$ and $n_{D,2k}$ are additive white Gaussian noise (AWGN) samples having zero mean and unit variance. We assume the channels are quasi--static within each slot and the channel SNRs of the $S$--$R$ and the $R$--$D$ links are $\gamma_{S,2k-1}$ and $\gamma_{R,2k}$, respectively. In particular, $\gamma_{S,2k-1}=|h_{S,2k-1}|^2$ and $\gamma_{R,2k}=|h_{R,2k}|^2$. We assume $\gamma_{S,2k-1}$ and $\gamma_{R,2k}$ to be independent and identically distributed (i.i.d.) over the time slots. Furthermore, $\gamma_{S,2k-1}$ and $\gamma_{R,2k}$ are mutually independent but not necessarily identically distributed (i.n.d.). For future reference, we introduce the average SNRs of the $S$--$R$ and the $R$--$D$ links as $\bar\gamma_{S}$ and $\bar\gamma_{R}$, respectively.

\noindent \textbf{System Throughput:} When $x_{2k-1}$ is transmitted from $S$ with transmit power $P_{S,2k-1}$ during time slot $2k-1$,
\beqa \xi_{S,2k-1} \triangleq{{\log }_2}\left( {1 + {\gamma _{S,2k-1}}{P_{S,2k-1}}} \right)\eeqa
bits of data can be transmitted error--free via the $S$--$R$ link. Similarly, when  $\hat x_{2k}$ is transmitted from $R$ with transmit power $P_{R,2k}$, \beqa \xi_{R,2k} \triangleq {{\log }_2}\left( {1 + {\gamma _{R,{2k}}}{P_{R,2k}}} \right)\eeqa
bits of data can be transmitted via the $R$--$D$ link. It is worth mentioning that during the $2k$th and the $(2k-1)$th time slots, $S$ and $R$, respectively, do not transmit any data, i.e., $P_{S,2k} = 0$ and $P_{R,2k-1} = 0$. We assume that $R$ ensures error--free detection by employing an appropriate error correction coding scheme and hence $\hat x_{2k} = x_{2k-1}$. Therefore, the end--to--end ($S$--$D$) system throughput is given by $\frac{1}{2}\min \{\xi_{S,2k-1},\xi_{R,2k}\}$ bits/s/Hz where the factor $\frac{1}{2}$ is due to the half--duplex constraint.

\noindent \textbf{Battery Dynamics:} The energies stored in the batteries of $S$ and $R$ in time slot $k$ are denoted by $B_{S,k}$ and $B_{R,k}$, respectively. The transmit powers of $S$ and $R$ are limited by the battery energies,  i.e., $0 \le P_{S,2k-1} \le B_{S,2k-1}$ and $0 \le P_{R,2k} \le B_{R,2k}$.  We assume throughout this paper that the energy consumed by the internal circuitry of $S$ and $R$ is negligible compared to the transmit power \cite{paper_EH_2}. The energy harvester at $S$ collects $H_{S,m} \le B_{S,max}$ units of energy during the $m$th time slot, where $m \in\{1,2,\cdots,K\}$. Similarly, the energy harvester at $R$ collects $H_{R,m} \le B_{R,max}$ units of energy during the $m$th time slot. It is worth noting that energies are harvested during every time slot $m$ at $S$ and $R$ regardless of $P_{S,m}$ and $P_{R,m}$. Let $H_{S,E} \triangleq E\{H_{S,m}\}$ and $H_{R,E} \triangleq E\{H_{R,m}\}$ denote the average energy harvesting rate of $S$ and $R$ over the time slots, respectively. Here, $E\{\cdot\}$ denotes statistical expectation. Because of the spatial separation of $S$ and $R$, we assume $H_{S,m}$ and $H_{R,m}$ are independent of each other and i.i.d. over the time slots. Similar to \cite{paper_EH_2}, we assume the stored energies at $S$ and $R$ increase and decrease linearly provided the maximum storage capacities, $B_{S,max}$ and $B_{R,max}$, are not exceeded, i.e.,
\begin{align}
& B_{S,m+1} = {\rm min} \{ (B_{S,m} - P_{S,m} + H_{S,m}), B_{S,max}  \},  \,\,\,\forall m \label{eq_new_1_pre}\\
& B_{R,m+1} = {\rm min} \{ (B_{R,m} - P_{R,m} + H_{R,m}), B_{R,max}  \},  \,\,\,\forall m
\label{eq_new_1}.
\end{align}
Furthermore, $B_{S,1}=H_{S,0} \ge 0$ and $B_{R,1}=H_{R,0} \ge 0$, respectively, denote the available energies at $S$ and $R$ before transmission starts.

\subsection{Buffer--Aided Adaptive Link Selection}

\noindent\textbf{Signal Model:} For buffer--aided link adaptive relaying, the relay $R$ is equipped with a buffer in which it can temporarily store the packets received from $S$. In this approach, one of the nodes decides whether $S$ or $R$ should transmit in a given time slot, $k \in \{1,2,\cdots,K\}$, based on the CSI of the $S$--$R$ and the $R$--$D$ links \cite{paper_EH_16}. Therefore, unlike conventional relaying, in any time slot $k$, $S$ or $R$ can transmit packets. Let $d_k \in \{ 0, 1 \}$ denote a binary link selection variable, where $d_k = 0$ ($d_k = 1$) if the $S$--$R$ ($R$--$D$) link is selected for transmission. When $d_k=0$, the received packet at $R$ is given by
\beqa y_{R,k} = h_{S,k}x_{k} + n_{R,k} \label{eq_A_1_two}.\eeqa On the other hand, when $d_k=1$, the received packet at $D$ is given by \beqa y_{D,k} = h_{R,k} \hat x_{k} + n_{D,k} \label{eq_A_2_two}.\eeqa

\noindent\textbf{System Throughput:} When $d_k=0$, $S$ is selected for transmission and
\beqa \bar\xi_{S,k} \triangleq (1 - d_k) {{\log }_2}\left( {1 + {\gamma _{S,k}}{P_{S,k}}} \right)\eeqa bits of data can be transmitted error--free via the $S$--$R$ link. Hence, $R$ receives $\bar\xi_{S,k}$ data bits from $S$ and appends them to the queue in its buffer. Therefore, the number of bits in the buffer at $R$ at the end of the $k$th time slot is denoted as $Q_k$ and given by $Q_k = Q_{k-1} + \bar\xi_{S,k}$. However, when $d_k=1$, $R$ transmits and the number of bits transmitted via the $R$--$D$ link is given by \beqa \bar\xi_{R,k} \triangleq {\min}\{ d_k {{\log }_2}\left( {1 + {\gamma _{R,k}}{P_{R,k}}} \right), Q_{k-1}\}.\eeqa It is worth noting that the maximal number of bits that can be sent by $R$ is limited by the number of bits in the buffer and the instantaneous capacity of the $R$--$D$ link \cite{paper_EH_16}. The number of bits remaining in the buffer at the end of the $k$th time slot is given by $Q_k = Q_{k-1} - \bar\xi_{R,k}$. We assume that $S$ has always data to transmit and the buffer at $R$ has very large (possibly infinite) capacity to store them. Therefore, a total of $\sum\limits_{k=1}^{K} \bar\xi_{R,k}$ bits are transmitted from $S$ to $D$ during the entire transmission time.

\noindent\textbf{Battery Dynamics:} The battery dynamics for the link adaptive transmission protocol are identical to those for conventional relaying.

%%%%%%%%%%%%%
\section{Power Allocation for Conventional Relaying \label{s3}}
In this section, we propose an offline and several online power allocation schemes for the considered EH system with conventional relaying.
\subsection{Optimal Offline Power Allocation}
%\vspace*{-3mm}
Our objective is to maximize the total number of transmitted bits (from $S$ to $D$) delivered by a deadline of $K$ time slots over a fading channel assuming offline (prior) knowledge of the full CSI and the energy arrivals at $S$ and $R$ in each time slot. The resulting maximization problem is subject to a causality constraint on the harvested energy and the (maximum) storage constraint for the batteries at both $S$ and $R$.

The offline optimization problem for maximizing the throughput of the considered system for $K$ time slots can be formulated as follows:
\begin{align}
& \underset{\mathcal{T} \succeq 0 }{\max}
& & \sum\limits_{k=1}^{K/2}{}  \min \{ \xi_{S,2k-1},\xi_{R,2k} \} \label{eq1A}\\
& \text{s.t.}
& & \small{\sum\limits_{k=1}^{l} (P_{S,2k-1} + \lambda_{S,2k-1}) \leq \sum\limits_{k=0}^{2(l-1)} H_{S,k}, \,\,\, \forall l} \label{eq2A}\\
&&& \small{\sum\limits_{k=1}^{l} (P_{R,2k} + \lambda_{R,2k}) \leq \sum\limits_{k=0}^{2l-1} H_{R,k}, \,\,\, \forall l} \label{eq3A}\\
&&& \small{\sum\limits_{k=0}^{2m-1} H_{S,k} - \sum\limits_{k=1}^{m} (P_{S,2k-1} + \lambda_{S,2k-1}) \leq B_{S,max}, \,\, \forall m } \label{eq4A_nw}\\
&&& {\small{\sum\limits_{k=0}^{2m} H_{R,k} - \sum\limits_{k=1}^{m} (P_{R,2k} + \lambda_{R,2k}) \leq B_{R,max}, \,\, \forall m}}  \label{eq5A}\\
&&& \gamma_{S,2k-1}P_{S,2k-1} = \gamma_{R,2k}P_{R,2k}, \,\,\, \forall k  \label{eq6A},
%&&& P_{S,2k} = 0, \,\,\, \forall k  \label{eq7A}\\
%&&& \lambda_{S,2k} = 0, \,\,\, \forall k  \label{eq8A}\\
%&&& P_{R,2k-1} = 0, \,\,\, \forall k  \label{eq9A}\\
%&&& P_{R,2k-1} = 0, \,\,\, \forall k  \label{eq10A},
\end{align}
where $\mathcal{T} \triangleq \{P_{S,2k-1},\, P_{R,2k},\, \lambda_{S,2k-1},\, \lambda_{R,2k} | k \in \{ 1,2,\cdots, K/2\} \}$. Also, $\forall l$, $\forall m$, and $\forall k$ stand for $l \in \{ 1,2,\cdots,K/2 \}$, $m \in \{ 1,2,\cdots,K/2-1 \}$, and $k \in \{ 1,2, \cdots, K/2 \} $, respectively. The slack variables $\lambda_{S,2k-1}$ and $\lambda_{R,2k}$ ensure that constraints (\ref{eq4A_nw})--(\ref{eq6A}) can be met for all realizations of $\gamma_{S,k}$, $\gamma_{R,k}$, $H_{S,k}$, and $H_{R,k}$. In particular, these slack variables represent the power (possibly) wasted in each transmission interval\footnote{For example, in case of small $B_{S,max}$ and $B_{R,max}$ and large $H_{S,E}$ and $H_{R,E}$, constraints (\ref{eq4A_nw}) and (\ref{eq5A}) can become infeasible if $\lambda_{S,2k-1}$ and $\lambda_{R,2k}$ are not introduced. $\lambda_{S,2k-1}$ and $\lambda_{R,2k}$ in the constraints avoid infeasibility of the problem and represent the amounts of energy that cannot be used (by $S$ and $R$) in each time slot.}. Constraints (\ref{eq2A}) and (\ref{eq3A}) stem from the causality requirement on the energy harvested at $S$ and $R$, respectively. Moreover, (\ref{eq4A_nw}) and (\ref{eq5A}) ensure that the harvested energy does not exceed the limited storage capacity of the batteries at $S$ and $R$, respectively. Constraint (\ref{eq6A}) ensures that the amount of information transmitted from $S$ to $R$ is identical to that transmitted from $R$ to $D$ so as to avoid data loss at $R$. Constraint (\ref{eq6A}) is required since we assume individual power constraints for $S$ and $R$. This is a reasonable assumption since $S$ and $R$ have independent power supplies.

Using (\ref{eq6A}) in (\ref{eq1A})--(\ref{eq5A}), the considered offline optimization problem can be rewritten as:
\begin{align}
& \underset{\mathcal{T}' \succeq 0}{\max}
& & \sum\limits_{k=1}^{K/2}  \xi_{S,2k-1} \label{eq1C}\\
& \text{s.t.}
%& & \sum\limits_{k=1}^{l} P_{S,k} \leq \sum\limits_{k=0}^{l-1} H_{S,k}, \,\,\, \forall l \label{eq2C}\\
& & \small{\sum\limits_{k=1}^{l} \left( \frac{\gamma_{S,2k-1} P_{S,2k-1}}{\gamma_{R,2k}} + \lambda_{R,2k} \right) \leq \sum\limits_{k=0}^{2l-1} H_{R,k}, \,\,\, \forall l} \label{eq3C}\\
%&&& \sum\limits_{k=1}^{l} H_{S,k} - \sum\limits_{k=1}^{l} P_{S,k} \leq B_{S,max}, \,\,\, \forall l  \label{eq4C}\\
&&& \small{\sum\limits_{k=0}^{2m} H_{R,k} - \sum\limits_{k=1}^{m} \left( \frac{\gamma_{S,2k-1} P_{S,2k-1}}{\gamma_{R,2k}} +  \lambda_{R,2k} \right) \leq B_{R,max} , \,\,\, \forall m } .\label{eq5C}\\
&&& {\rm Constraints  \,\,(\ref{eq2A}) \,\,and\,\, (\ref{eq4A_nw})} ,\nonumber
\end{align}
where $\mathcal{T}' \triangleq \mathcal{T} \setminus \{ P_{R,2k} | \forall k\}$. The problem in (\ref{eq1C})--(\ref{eq5C}) with (\ref{eq2A}) and (\ref{eq4A_nw}) forms a convex optimization problem and the optimum solution can be obtained either in closed form by using the Karush–-Kuhn–-Tucker (KKT) conditions\footnote{It can easily be shown that optimization problem (\ref{eq1C})--(\ref{eq5C}) with (\ref{eq2A}) and (\ref{eq4A_nw}) has a water--filling solution. Due to space limitation, we omit the solution here.} or by using any standard technique for solving convex optimization problems \cite{paper44}, \cite{paper30}. Let $P_{S,2k-1}^*$ denote the optimum solution of the considered optimization problem. The optimum $P_{R,2k}$ can be obtained as
\beqa
P_{R,2k}^{*} = \frac{\gamma_{S,2k-1}P_{S,2k-1}^{*}}{\gamma_{R,2k}}
\label{eq_nw_4}.
\eeqa

\subsection{Optimal Online Power Allocation by DP}
In practice, only causal information about channels and harvested energies is available for power allocation. Therefore, the offline power allocation scheme is not readily applicable as, at a given time slot, the future CSI and the upcoming harvested energy are not known in advance. We propose to employ a stochastic DP approach for optimum online power allocation \cite{paper_EH_2, paper45}.

Let $\cvec_{2k-1,2k} \triangleq (\gamma_{S,2k-1},\gamma_{R,2k},(H_{S,2(k-1)}+H_{S,2k-3}),(H_{R,2k-1}+H_{R,2(k-1)}),B_{S,2k-1},B_{R,2k})$ denote the state for time slots $2k-1$ and $2k$. We note that $H_{S,k} = 0$ for $k<0$. Our aim is to maximize the total throughput over $K$ time slots. We assume the initial state $\cvec_{1,2} = (\gamma_{S,1},\gamma_{R,2},H_{S,0},(H_{R,0}+H_{R,1}),B_{S,1},B_{R,2})$ is always known. We define a policy $p = \{ (P_{S,2k-1}\\(\cvec_{2k-1,2k}),P_{R,2k}(\cvec_{2k-1,2k})), \forall \cvec_{2k-1,2k}, k = 1,2,\cdots,K/2\}$ as feasible if the energy harvesting constraints $0 \le P_{S,2k-1}(\cvec_{2k-1,2k}) \le B_{S,2k-1}$ and $0 \le P_{R,2k}(\cvec_{2k-1,2k}) \le B_{R,2k}$ are satisfied for all $k$. Hence, the objective function to be maximized can be reformulated as \cite{paper_EH_2}
\beqa R(p) =  \sum\limits_{k=1}^{K/2}{}  E \{\min \{ \xi_{S,2k-1}',\xi_{R,2k}' \} | \cvec_{1,2}, p\}, \label{eq_new_last_1}
\eeqa
where we defined $\xi_{S,2k-1}' \triangleq {{\log }_2}\left( {1 + {\gamma _{S,2k-1}}{P_{S,2k-1}(\cvec_{2k-1,2k})}} \right)$ and $ \xi_{R,2k}' \triangleq \log_2 ( 1 + \gamma_{R,2k} P_{R,2k}$ {$ (\cvec_{2k-1,2k}))$}. \normalsize The expectation is with respect to the SNRs of the channels and the harvested energies. In particular, for a given $\cvec_{1,2}$, the maximum throughput can be obtained as \beqa R^* = \underset{p\in \mathcal{P}}{\max} \,\, R(p) ,\eeqa  where $\mathcal{P}$ denotes the space of all feasible policies.

The maximum throughput during time slots $2k-1$ and $2k$ is denoted by $J_{2k-1,2k}(B_{S,2k-1},B_{R,2k})$. For a given $\cvec_{1,2}$, the maximum throughput, $J_{1,2}(B_{S,1},B_{R,2})$, can be recursively obtained from $J_{K-1,K}(B_{S,K-1},B_{R,K})$, $J_{K-3,K-2}(B_{S,K-3},B_{R,K-2})$, $\cdots$, $J_{3,4}(B_{S,3},B_{R,4})$ \cite{paper_EH_2}. For the last two time slots $K-1$ and $K$, we have
\beqa
J_{K-1,K}(B_{S,K-1},B_{R,K}) = \mathop {\max }\limits_{\scriptstyle\,\,\,\,\,\,\,0 \le {P_{S,K-1}} \le {B_{S,K-1}}\hfill\atop
{\scriptstyle\,\,\,\,\,\,\,\,\,\,0 \le {P_{R,K}} \le {B_{R,K}}\hfill\atop
\scriptstyle{\gamma _{S,K-1}}{P_{S,K-1}} = {\gamma _{R,K}}{P_{R,K}}\hfill}} \frac{1}{2}\min \left\{ {\xi_{S,K-1},\xi_{R,K}} \right\}
\label{eq4A_DP}
\eeqa
and for time slots $2k-1$ and $2k$, we obtain
\beqa
J_{2k-1,2k}(B_{S,2k-1},B_{R,2k}) = \mathop {\max }\limits_{\scriptstyle\,\,\,\,\,\,\,0 \le {P_{S,2k-1}} \le {B_{S,{2k-1}}}\hfill\atop
{\scriptstyle\,\,\,\,\,\,\,0 \le {P_{R,2k}} \le {B_{R,{2k}}}\hfill\atop
\scriptstyle{\gamma _{S,{2k-1}}}{P_{S,2k-1}} = {\gamma _{R,{2k}}}{P_{R,2k}}\hfill}} \frac{1}{2}\min \left\{ {\xi_{S,2k-1},\xi_{R,2k}} \right\} \nonumber \\
&&\hspace*{-6.1cm} + \bar J_{2k+1,2k+2}(B_{S,2k-1}-P_{S,2k-1} , B_{R,2k}-P_{R,2k}),
\label{eq4B_DP}
\eeqa
where $\bar J_{2k+1,2k+2}(B_{S,2k+1}' , B_{R,2k+2}') = $
\beqa
{E}_{\tilde\gamma_{S,2k+1},\tilde\gamma_{R,2k+2},\tilde H_{S,2k-1},\tilde H_{R,2k}} \{ J_{2k+1,2k+2}(\min \{B_{S,2k+1}'+\tilde H_{S,2k-1},B_{S,max}\},\nonumber \\
&&\hspace*{-5.6cm}\min \{B_{R,2k+2}'+\tilde H_{R,2k},B_{R,max}\}) \}
\label{eq4C_DP}.
\eeqa
Here, $B_{S,2k+1}' \triangleq B_{S,2k-1} - P_{S,2k-1}$, $B_{R,2k+2}' \triangleq B_{R,2k} - P_{R,2k}$, $\tilde \gamma_{S,2k+1}$ ($\tilde \gamma_{R,2k+2}$) represents the SNR of the $S$--$R$ ($R$--$D$) link in the $(2k+1)$th ($(2k+2)$th) slot given the SNR $\gamma_{S,2k-1}$ ($\gamma_{R,2k}$) in the $(2k-1)$th ($2k$th) slot, and $\tilde H_{S,2k-1}$ ($\tilde H_{R,2k}$) denotes the harvested energy at $S$ ($R$) in the $(2k-1)$th ($2k$th) slot given the harvested energy $H_{S,2k-2}$ ($H_{R,2k-1}$) in the $(2k-2)$th ($(2k-1)$th) slot. It can be shown that the cost functions in (\ref{eq4A_DP}) and (\ref{eq4B_DP}) are concave in $P_{S,2k-1}$ and $P_{R,2k}$. Thus, (\ref{eq4A_DP}) and (\ref{eq4B_DP}) are convex optimization problems and can be solved very efficiently \cite{paper44}. Further simplification of (\ref{eq4A_DP}) yields
\beqa
J_{K-1,K}(B_{S,K-1},B_{R,K}) = \frac{1}{2} \log_2 \left( 1 + \gamma_{S,K-1}\rho_{K-1} \right)
\label{eq_new_5A_DP},
\eeqa
where $\rho_{K-1} = \min \left\{ B_{S,K-1},  \gamma_{R,K}B_{R,K}/\gamma_{S,K-1} \right\}$. Therefore, $P_{S,K-1}^* = \min \{B_{S,K-1},\frac{\gamma_{R,K}B_{R,K}}{\gamma_{S,K-1}}\}$, and $P_{R,K}^*$ follows from (\ref{eq_nw_4}). Similarly, (\ref{eq4B_DP}) can be simplified as
\beqa
J_{2k-1,2k}(B_{S,2k-1},B_{R,2k}) \nonumber  = \mathop {\max }\limits_{\scriptstyle 0 \le {P_{S,{2k-1}}} \le \min\left\{{B_{S,2k-1}, \gamma_{R,2k}B_{R,2k}/\gamma_{S,2k-1} }\right\}\hfill\atop}  \frac{1}{2}\xi_{S,2k-1} \nonumber \\
&&\hspace*{-9.6cm} + \bar J_{2k+1,2k+2}(B_{S,2k-1}-P_{S,2k-1} , B_{R,2k}- \gamma_{S,2k-1}P_{S,2k-1}/\gamma_{R,2k} )
\label{eq_new_6A_DP}.
\eeqa
Using (\ref{eq_new_5A_DP}) and (\ref{eq_new_6A_DP}), $P_{S,2k-1}^*$ and $P_{R,2k}^*$, $k \in \{1,2,\cdots,K/2\}$, can be obtained for different possible values of $\gamma_{S,k}$, $\gamma_{R,k}$, $B_{S,k}$, and $B_{R,k}$ and can be stored in a look--up table. This is done before transmission starts. When transmission starts, for a given realization of $\gamma_{S,2k-1}$, $\gamma_{R,2k}$, $B_{S,2k-1}$, and $B_{R,2k}$ in time slots $2k-1$ and $2k$, the values of $P_{S,2k-1}^*$ and $P_{R,2k}^*$ corresponding to that realization are taken from the look--up table.

\subsection{Suboptimal Online Power Allocation}
In the proposed DP--based optimal online power allocation scheme, for a certain transmission time slot, we consider the average effect of all succeeding time slots, c.f. (\ref{eq4C_DP}). Due to the recursive nature of DP, the computational complexity of this approach increases alarmingly with increasing $K$. For this reason, in the following, we propose three different suboptimal online power allocation schemes, which perform close to the optimal DP approach but have reduced complexity.

\subsubsection{Suboptimal Simplified DP Power Allocation (``DP--${\rm I}_2$" and ``DP--${\rm I}_1$" Schemes)}
In this scheme, we use the average effect of only 2 (or 4) following time slots to allocate the transmit power in each time slot. In particular, we assume for the current time slot that all energies have to be spent over the following 2 (or 4) time slots. Moreover, in the last two time slots, either $S$ or $R$ uses up all of its stored energy. This scheme reduces the computational complexity at the expense of a performance degradation. We refer to the suboptimal DP schemes taking into account 4 and 2 time slots as ``DP--${\rm I}_2$" and ``DP--${\rm I}_1$", respectively.

\subsubsection{Suboptimal Harvesting Rate (HR) Assisted Power Allocation (``HR Assisted" Scheme)}
In this scheme, we constrain the transmit powers $P_{S,2k-1}$ and $P_{R,2k}$ by the average energy harvesting rates $H_{S,E}$ and $H_{R,E}$, respectively.  This scheme is referred to as ``HR Assisted" power allocation. For a given time slot $k \in \{1,2,\cdots,K/2-1\}$, the resulting optimization problem can be stated as:
\begin{align}
& \underset{P_{S,2k-1} \geq 0}{\max}
&&   \xi_{S,2k-1} \label{eq1SC}\\
& \text{s.t.}
& & P_{S,2k-1} \leq B_{S,2k-1},\,\, P_{S,2k-1} \leq H_{S,E}, \label{eq2SC}\\
%&&& P_{S,2k-1} \leq H_{S,E}, \label{eq3SC}\\
&&& P_{S,2k-1} \leq \frac{\gamma_{R,2k}B_{R,2k}}{\gamma_{S,2k-1}},\,\, P_{S,2k-1} \leq \frac{\gamma_{R,2k}H_{R,E}}{\gamma_{S,2k-1}}  \label{eq4SC} %\\
%&&& P_{S,2k-1} \leq \frac{\gamma_{R,2k}H_{R,E}}{\gamma_{S,2k-1}}  .\label{eq5SC}
\end{align}
The optimum solution for this optimization problem is given by  \beqa P_{S,2k-1}^* = \min \{B_{S,2k-1}, H_{S,E}, \frac{\gamma_{R,2k}B_{R,2k}}{\gamma_{S,2k-1}}, \frac{\gamma_{R,2k}H_{R,E}}{\gamma_{S,2k-1}}\},\eeqa and $P_{R,2k}^*$ is obtained from (\ref{eq_nw_4}). For the $K$th time slot, we ensure that either $S$ or $R$ use up all of their stored energy.

\subsubsection{Suboptimal Naive Power Allocation (``Naive" Scheme)}
In this suboptimal ``naive" approach, for each time slot, only the stored energies at hand determine the transmit power, i.e., this approach does not take into account the effect of the following time slots, and the transmitting node ($S$ or $R$) uses up all of its available energy in each transmission interval. To be specific, for a particular time slot $k \in \{1,2,\cdots,K/2\}$, $P_{S,2k-1}^* = \min \{B_{S,2k-1},\frac{\gamma_{R,2k}B_{R,2k}}{\gamma_{S,2k-1}}\}$, and $P_{R,2k}^*$ follows directly from (\ref{eq_nw_4}).

\subsection{Complexity}
In the offline and the optimal online power allocation schemes, we solve convex optimization problems where the number of constraints is a function of $K$. The required computational complexity to solve a convex optimization problem depends on the method used (e.g. bisection method, interior--point--method, etc.) and is polynomial in the size of the problem \cite{paper44}. Therefore, the worst--case computational complexity of the offline power allocation scheme for conventional relaying is polynomial in the number of time slots $K$ \cite{paper44}. We observe from (\ref{eq4B_DP}) and (\ref{eq4C_DP}) that, in the optimal DP based online power allocation scheme, the optimal transmit powers of $S$ and $R$ at a given time slot depend on the average data rate of the next time slot. We consider different realizations of channel SNRs and harvested energies for the next time slot to obtain the average data rate. Furthermore, the data rates corresponding to each of the realizations of channel SNRs and harvested energies in the next time slot are also functions of the average date rates in the second next time slot. This dependency spans up to the last time slot and therefore, the complexity of the optimal online power allocation scheme increases exponentially with $K$. The less complex versions of DP, i.e., DP--${\rm I}_2$ and DP--${\rm I}_1$ have linear complexities in $K$. Moreover, the naive and the HR assisted suboptimal online schemes have linear complexities in $K$. Note that we compare the complexity of the proposed offline and online power allocation schemes in terms of the required average execution time in the simulation results section.

%%%%%%%%%%%%%
\section{Power Allocation for Buffer--Aided Adaptive Link Selection\label{s4}}
In this section, we propose offline and online power allocation schemes for EH systems with buffer--aided adaptive link selection.
\subsection{Offline Power Allocation}
Like for conventional relaying, our goal is to maximize the total number of transmitted bits (from $S$ to $D$) delivered by a deadline of $K$ time slots for the link adaptive transmission protocol. The offline (prior) information about the full CSI and the energy arrivals at $S$ and $R$ in each time slot are assumed to be known in advance. The offline optimization problem for the link adaptive transmission protocol can be formulated as follows:
\begin{align}
& \underset{\mathcal{T} \geq 0, \, \{d_k|\forall k\}}{\text{max}}
& & \sum\limits_{k=1}^{K} d_k \log _2(1 + \gamma_{R,k}P_{R,k})  \label{eq1A_buff}\\
& \text{s.t.}
& & \sum\limits_{k=1}^{q} ((1 - d_k)P_{S,k} + \lambda_{S,k}) \leq \sum\limits_{k=0}^{q-1} H_{S,k}, \,\,\, \forall q \label{eq2A_buff}\\
&&& \sum\limits_{k=1}^{q} (d_k P_{R,k} + \lambda_{R,k}) \leq \sum\limits_{k=0}^{q-1} H_{R,k}, \,\,\, \forall q \label{eq3A_buff}\\
&&& \sum\limits_{k=0}^{v} H_{S,k} - \sum\limits_{k=1}^{v} ((1 - d_k) P_{S,k} + \lambda_{S,k}) \leq B_{S,max}, \,\,\, \forall v  \label{eq4A_buff}\\
&&& \sum\limits_{k=1}^{v} H_{R,k} - \sum\limits_{k=1}^{v} (d_k P_{R,k} + \lambda_{R,k}) \leq B_{R,max}, \,\,\, \forall v  \label{eq5A_buff}\\
&&& \sum\limits_{k=1}^{q} (1 - d_k)\log_2\left( 1 + \gamma_{S,k}P_{S,k} \right) \geq \sum\limits_{k=1}^{q} d_k\log_2\left( 1 + \gamma_{R,k}P_{R,k} \right), \,\,\, \forall q \label{eq6A_buff}\\
&&& d_k(1 - d_k) = 0, \,\,\, \forall k,  \label{eq7A_buff}
\end{align}
where $\forall q$, $\forall k$, and $\forall v$ stand for $q \in \{1,2,\cdots,K\}$, $k \in \{1,2,\cdots,K\}$, and $v \in \{1,2,\cdots,K-1\}$, respectively. Like for conventional relaying, constraints (\ref{eq2A_buff})--(\ref{eq5A_buff}) ensure the energy causality and limited energy conditions for buffer--aided link adaptive relaying. Constraint (\ref{eq6A_buff}) ensures that $R$ cannot transmit more bits than it has in its buffer. Moreover, (\ref{eq7A_buff}) ensures that $d_k$ can only be 0 or 1, i.e., either $S$ or $R$ transmits in a given time slot, $k \in \{ 1,2,\cdots,K \}$. We note that, in this optimization problem, although we are maximizing the throughput of the $R$--$D$ link, using constraint (\ref{eq6A_buff}) incorporates the effect of the throughput of the $S$--$R$ link. As $\xi_{S,k}$ and $\xi_{R,k}$ are increasing functions of $P_{S,k}$ and $P_{R,k}$, respectively, the optimization problem in (\ref{eq1A_buff})--(\ref{eq7A_buff}) can be restated as follows:
\begin{align}
& \underset{\mathcal{T}'' \geq 0, \, \{d_k|\forall k\}}{\text{max}}
& & \sum\limits_{k=1}^{K} d_k \xi_{R,k}  \label{eq1A_buff2}\\
& \text{s.t.}
& & \sum\limits_{k=1}^{q} \left( \frac{(1 - d_k)(2^{\xi_{S,k}}-1)}{\gamma_{S,k}} + \lambda_{S,k} \right) \leq \sum\limits_{k=0}^{q-1} H_{S,k}, \,\,\, \forall q \label{eq2A_buff2}\\
&&& \sum\limits_{k=1}^{q} \left( \frac{d_k (2^{\xi_{R,k}}-1)}{\gamma_{R,k}} + \lambda_{R,k} \right) \leq \sum\limits_{k=0}^{q-1} H_{R,k}, \,\,\, \forall q \label{eq3A_buff2}\\
&&& \sum\limits_{k=0}^{v} H_{S,k} - \sum\limits_{k=1}^{v}  \left( \frac{(1 - d_k)(2^{\xi_{S,k}}-1)}{\gamma_{S,k}} + \lambda_{S,k} \right) \leq B_{S,max}, \,\,\, \forall v  \label{eq4A_buff2}\\
&&& \sum\limits_{k=1}^{v} H_{R,k} - \sum\limits_{k=1}^{v}  \left( \frac{d_k(2^{\xi_{R,k}}-1)}{\gamma_{R,k}} + \lambda_{R,k} \right) \leq B_{R,max}, \,\,\, \forall v  \label{eq5A_buff2}\\
&&& \sum\limits_{k=1}^{q} (1 - d_k)\xi_{S,k} \geq \sum\limits_{k=1}^{q} d_k \xi_{R,k}, \,\,\, \forall q \label{eq6A_buff2}\\
&&& d_k(1 - d_k) = 0, \,\,\, \forall k,  \label{eq7A_buff2}
\end{align}
where $\mathcal{T}'' = \{\xi_{S,k} ,\, \xi_{R,k} ,\, \lambda_{S,k} ,\, \lambda_{R,k}| k \in \{1,2,\cdots,K \}\}$. The problem in (\ref{eq1A_buff2})--(\ref{eq7A_buff2}) is a non--convex MINLP due to the binary variables $d_k$ and the non--convex and non--linear constraints (\ref{eq2A_buff2})--(\ref{eq7A_buff2}). In the following, we propose optimal methods to solve the buffer--aided link adaptive offline optimization problem.

%\subsubsection{Optimal solution}
%In this subsection, we discuss two methods to solve our offline optimization problem optimally.\\

\subsubsection{Exhaustive Search}
For given $d_k$, $k \in \{1,2,\cdots,K\}$, the optimization problem in (\ref{eq1A_buff2})--(\ref{eq7A_buff2}) is convex. Therefore, we can optimize $\xi_{S,k}$ and $\xi_{R,k}$ for given $d_k \in \{0,1\}$ very efficiently. In this method, we optimize $\xi_{S,k}$ and $\xi_{R,k}$ for all possible combinations of $d_k$, $k \in \{1,2,\cdots,K\}$, and select from all the solutions that combination of $d_k$, $k \in \{1,2,\cdots,K\}$, which maximizes the cost function. This exhaustive search method provides the global optimal solution but with an exponential complexity. For instance, for $K$ time slots we have $2^{K-2}$ feasible combinations of $d_k$ and hence to optimize $\xi_{S,k}$ and $\xi_{R,k}$, we need to solve $2^{K-2}$ optimization problems. Therefore, in practice, this approach cannot be adopted in general, especially for large $K$. However, the exhaustive search scheme can be effective for small $K$.

\subsubsection{Spatial Branch-and-Bound}
As mentioned before, our problem is a non--convex MINLP. One of the recent advances in (globally) solving MINLP problems is the sBB method \cite{paper46,paper47}. The sBB method sequentially solves subproblems of problem (\ref{eq1A_buff2})--(\ref{eq7A_buff2}). These subproblems are obtained by partitioning the original solution space. For each subproblem, the sBB method relies on the generation of rigorous lower and upper bounds of the problem over any given variable sub--domain. The feasible lower bounds are chosen to be the local minimizers of the (sub)problems whereas the upper bounds are obtained from convex relaxations. Interestingly, MINLP problems can be solved by using the widely available open source solver Couenne \cite{paper49,paper47}. Couenne provides the global optimal solution for both convex and non--convex MINLP problems. It implements linearization, bound reduction, and branching methods within a branch and bound framework. For global convergence and complexity issues of sBB and therefore Couenne, please refer to \cite{paper47}.

\subsection{Online Power Allocation}
As offline power allocation requires non--causal knowledge of CSI and harvested energies, we also propose two sub--optimal online power allocation schemes, which require only causal knowledge of CSI and harvested energy.
\subsubsection{Online Power Allocation Based on Average Data and Average Harvesting Rates}

Unlike for conventional relaying, for the link adaptive protocol, the use of DP for online power allocation is not possible because of the link selection. For this reason, for the online power allocation scheme, at first we formulate an optimization problem which is based on the average data rate, the average energy causality constraints at $S$ and $R$, and the average buffering constraint for $K \to \infty$. We assume that the batteries at $S$ and $R$ have unlimited capacities. Considering the above mentioned assumptions, the optimization problem for online power allocation can be formulated as:

\begin{align}
& \underset{\{ P_{S,k} \geq 0, \, P_{R,k} \geq 0, \, d_k | \forall k\} }{\text{max}}
& & \frac{1}{K} \sum\limits_{k=1}^{K} d_k \log _2(1 + \gamma_{R,k}P_{R,k})  \label{eqB1A}\\
& \text{s.t.}
& & \frac{1}{K}\sum\limits_{k=1}^{K} (1 - d_k) P_{S,k} \leq \frac{1}{K}\sum\limits_{k=0}^{K-1} H_{S,k} \label{eqB2A}\\
&&& \frac{1}{K}\sum\limits_{k=1}^{K} d_k P_{R,k} \leq \frac{1}{K}\sum\limits_{k=0}^{K-1} H_{R,k} \label{eqB3A}\\
&&& \frac{1}{K}\sum\limits_{k=1}^{K} (1 - d_k)\log_2\left( 1 + \gamma_{S,k}P_{S,k} \right) = \frac{1}{K}\sum\limits_{k=1}^{K} d_k\log_2\left( 1 + \gamma_{R,k}P_{R,k} \right) \label{eqB4A}\\
&&& \frac{1}{K} d_k(1 - d_k) = 0, \,\,\, \forall k  \label{eqB5A}
\end{align}

\noindent The Lagrangian of (\ref{eqB1A})--(\ref{eqB5A}) is given by
\beqa
\mathcal{L} = \frac{1}{K}\sum\limits_{k=1}^{K}d_k \log_2(1+\gamma_{R,k}P_{R,k}) - \frac{\lambda_S}{K}\left( \sum\limits_{k=1}^{K} (1-d_k)P_{S,k} - H_{S,k-1} \right) \nonumber\\
&&\hspace*{-9.6cm} - \frac{\lambda_R}{K}\left( \sum\limits_{k=1}^{K} d_k P_{R,k} - H_{R,k-1} \right) - \frac{1}{K}\sum\limits_{k=1}^{K}\beta_{k}d_k(1-d_k) \nonumber\\
&&\hspace*{-9.6cm} - \frac{\mu}{K} \left( \sum\limits_{k=1}^K d_k \log_2(1+\gamma_{R,k}P_{R,k}) - (1-d_k)\log_2(1+\gamma_{S,k}P_{S,k})  \right),
\label{eqB2}
\eeqa
where $\lambda_S$, $\lambda_R$, $\mu$, and $\beta_k$ are Lagrange multipliers. Differentiating (\ref{eqB2}) with respect to $P_{S,k}$, $P_{R,k}$, and $d_k$ and equating each of the differentiated expressions to zero leads to the following optimum values of $P_{S,k}$, $P_{R,k}$, and $d_k$:
\begin{equation}
P_{S,k}^* = \left\{ \begin{array}{l}
\frac{\rho}{\nu_S} - \frac{1}{\gamma_{S,k}}, \,\,\,\,\, {\rm if} \,\, \gamma_{S,k} > \frac{\nu_S}{\rho} \,\, {\rm AND} \,\, d_k=0 \label{eqB3},\\
0, \,\,\, {\rm otherwise},
\end{array} \right.
\end{equation}
\begin{equation}
P_{R,k}^* = \left\{ \begin{array}{l}
\frac{1}{\nu_R} - \frac{1}{\gamma_{R,k}}, \,\,\,\,\, {\rm if} \,\, \gamma_{R,k} > {\nu_R} \,\, {\rm AND} \,\, d_k=1, \label{eqB4} \\
0, \,\,\, {\rm otherwise},
\end{array} \right.
\end{equation}
\begin{equation}
d_k^* = \left \{ \begin{array}{l}
1, \,\,\,\,\, {\rm if}\,\, \left( \mathcal{C}_{R} > \mathcal{C}_{S} \right)  \,\, {\rm OR} \,\, \left( \gamma_{R,k} > \nu_R \,\, {\rm AND} \,\, \gamma_{S,k} < \frac{\nu_S}{\rho} \right) , \label{eqB6_pre}\\
0, \,\,\,\,\, {\rm if}\,\, \left(\mathcal{C}_{R} < \mathcal{C}_{S}\right) {\rm OR} \,\, \left( \gamma_{R,k} < \nu_R \,\, {\rm AND} \,\, \gamma_{S,k} > \frac{\nu_S}{\rho} \right),
\end{array} \right.
\end{equation}
where $\mathcal{C}_{R} = \ln \left( \frac{\gamma_{R,k}}{\nu_R}\right) + \frac{\nu_R}{\gamma_{R,k}} - 1$, $\mathcal{C}_{S} = \rho \ln\left( \frac{\rho\gamma_{S,k}}{\nu_S} \right) + \frac{\nu_S}{\gamma_{S,k}} - \rho$, $\rho=\mu/(1-\mu)$, $\nu_S=\lambda_S\ln(2)/(1-\mu)$, and $\nu_R=\lambda_R\ln(2)/(1-\mu)$. We observe that the optimal $P_{S,k}$, $P_{R,k}$, and $d_k$ depend on the instantaneous channel SNRs and Lagrange multipliers. The Lagrange multipliers can be solved efficiently, as shown in the next part of this section, without requiring any non--causal knowledge. Therefore, the optimal $P_{S,k}$, $P_{R,k}$, and $d_k$ are readily applicable in the real--time (online) environment with low implementation complexity.

\noindent\textbf{Finding the Lagrange Multipliers:}
\noindent Combining (\ref{eqB3})--(\ref{eqB6_pre}) and (\ref{eqB2A})--(\ref{eqB4A}) yields the following conditions for $K \to \infty$:
\begin{align}
& \int\limits_{0}^{\nu_R}\left[ \int\limits_{\frac{\nu_S}{\rho}}^{\infty} \left( \frac{\rho}{\nu_S}-\frac{1}{\gamma_{S,k}} \right)f_{\gamma_{S,k}}(\gamma_{S,k}) {\rm d}\gamma_{S,k} \right] f_{\gamma_{R,k}}(\gamma_{R,k}) {\rm d}\gamma_{R,k} \nonumber \\
&&\hspace*{-6.6cm} + \int\limits_{\nu_R}^{\infty}\left[ \int\limits_{L_1}^{\infty} \left( \frac{\rho}{\nu_S}-\frac{1}{\gamma_{S,k}} \right)f_{\gamma_{S,k}}(\gamma_{S,k}) {\rm d}\gamma_{S,k} \right] f_{\gamma_{R,k}}(\gamma_{R,k}) {\rm d}\gamma_{R,k} = H_{S,E},\label{eqB7}
\end{align}
\begin{align}
& \int\limits_{0}^{\frac{\nu_S}{\rho}}\left[ \int\limits_{{\nu_R}}^{\infty} \left( \frac{1}{\nu_R}-\frac{1}{\gamma_{R,k}} \right)f_{\gamma_{R,k}}(\gamma_{R,k}) {\rm d}\gamma_{R,k} \right] f_{\gamma_{S,k}}(\gamma_{S,k}) {\rm d}\gamma_{S,k} \nonumber \\
&&\hspace*{-6.6cm} + \int\limits_{\frac{\nu_S}{\rho}}^{\infty}\left[ \int\limits_{L_2}^{\infty} \left( \frac{1}{\nu_R}-\frac{1}{\gamma_{R,k}} \right)f_{\gamma_{R,k}}(\gamma_{R,k}) {\rm d}\gamma_{R,k} \right] f_{\gamma_{S,k}}(\gamma_{S,k}) {\rm d}\gamma_{S,k} = H_{R,E}, \label{eqB8}
\end{align}
\begin{align}
& \int\limits_{0}^{\nu_R}\left[ \int\limits_{\frac{\nu_S}{\rho}}^{\infty}  \log_2\left( \frac{\rho\gamma_{S,k}}{\nu_S} \right) f_{\gamma_{S,k}}(\gamma_{S,k}) {\rm d}\gamma_{S,k} \right] f_{\gamma_{R,k}}(\gamma_{R,k}) {\rm d}\gamma_{R,k} \nonumber \\
&&\hspace*{-6.6cm} + \int\limits_{\nu_R}^{\infty}\left[ \int\limits_{{L_1}}^{\infty}  \log_2\left( \frac{\rho\gamma_{S,k}}{\nu_S} \right) f_{\gamma_{S,k}}(\gamma_{S,k}) {\rm d}\gamma_{S,k} \right] f_{\gamma_{R,k}}(\gamma_{R,k}) {\rm d}\gamma_{R,k} \nonumber \\
&\hspace*{+0.0cm} = \int\limits_{0}^{\frac{\nu_S}{\rho}}\left[ \int\limits_{{\nu_R}}^{\infty}  \log_2\left( \frac{\gamma_{R,k}}{\nu_R} \right) f_{\gamma_{R,k}}(\gamma_{R,k}) {\rm d}\gamma_{R,k} \right] f_{\gamma_{S,k}}(\gamma_{S,k}) {\rm d}\gamma_{S,k} \nonumber \\
&&\hspace*{-6.6cm} + \int\limits_{\frac{\nu_S}{\rho}}^{\infty}\left[ \int\limits_{L_2}^{\infty}  \log_2\left( \frac{\gamma_{R,k}}{\nu_R} \right) f_{\gamma_{R,k}}(\gamma_{R,k}) {\rm d}\gamma_{R,k} \right] f_{\gamma_{S,k}}(\gamma_{S,k}) {\rm d}\gamma_{S,k} \label{eqB9},
\end{align}
where $L_1 = - \frac{\nu_S}{\rho {\rm W}\left( - e^{\frac{\gamma_{R,k}-\nu_R}{\rho\gamma_{R,k}}-1} \left( \frac{\nu_R}{\gamma_{R,k}} \right)^{\frac{1}{\rho}} \right)}$ and $L_2 = - \frac{\nu_R}{{\rm W} \left( -e^{\rho - \frac{\nu_S}{\gamma_{S,k}} - 1} \left( \frac{\nu_S}{\rho\gamma_{S,k}} \right)^{\rho} \right)}$. Here, $W(\cdot)$ is the Lambert W--function \cite{paper_EH_19}, and $f_{\gamma_{S,k}}(\gamma_{S,k})$ and $f_{\gamma_{R,k}}(\gamma_{R,k})$ denote the probability density functions (pdfs) of the $S$--$R$ and $R$--$D$ channel SNRs, respectively. For Rayleigh fading, $f_{\gamma_{S,k}}(\gamma_{S,k}) = ({1}/{\bar\gamma_{S}}) e^{-\gamma_{S,k}/\bar\gamma_{S}}$ and $f_{\gamma_{R,k}}(\gamma_{R,k}) = ({1}/{\bar\gamma_{R}}) e^{-\gamma_{R,k}/\bar\gamma_{R}}$. We need to solve (\ref{eqB7})--(\ref{eqB9}) to find the optimal $\nu_S$, $\nu_R$, and $\rho$. The solution can be obtained by using the built--in root--finding function in Mathematica. We note that the optimal $\nu_S$, $\nu_R$, and $\rho$ are computed offline before transmission starts. When transmission begins, $P_{S,k}^*$, $P_{R,k}^*$, and $d_{k}^*$ are calculated based on offline parameters $\rho$, $\nu_S$, and $\nu_R$ and online variables $\gamma_{S,k}$ and $\gamma_{R,k}$.

The solution of problem (\ref{eqB1A})--(\ref{eqB5A}) provides an upper bound for the practical case where the storage capacity of the batteries is limited. Moreover, the problem may yield $P_{R,k} \ne 0$ even if the buffer is empty at $R$. To avoid this undesirable behavior, we propose a practical but suboptimal online algorithm which is summarized in Algorithm 1. At first, we calculate $B_{S,k}$ and $B_{R,k}$ using (\ref{eq_new_1_pre}) and (\ref{eq_new_1}), respectively. We then calculate $P_{S,k}^*$, $P_{R,k}^*$, and $d_k^*$ from (\ref{eqB3})--(\ref{eqB6_pre}) and (\ref{eqB2A})--(\ref{eqB4A}). To ensure that $P_{S,k}^*$ and $P_{R,k}^*$ do not exceed the storage limits, we perform steps 6 to 8 and 11 to 13, respectively. Steps 9 and 17 keep track of the arrival of data bits into and the departure of data bits out of the buffer, respectively. Steps 14 to 16 are adopted to ensure that $R$ transmits only if there is data in the buffer.

%%%%%%%%%%%%%%%%%%%%%%%%%%%%%%%%%%%%%%%%%%%%%%%%%%%%%%%%%%%%%%%%%

\subsubsection{Naive Online Power Allocation}
In the suboptimal naive power allocation scheme for link adaptive relaying, at each time slot, $k$, $S$ and $R$ consider the amount of energy stored in their batteries as their transmit powers. Based on the transmit powers, $S$ and $R$ compute their capacities. Note that the buffer status should be taken into account in the computation of the capacity of $R$.  The $S$--$R$ ($R$--$D$) link is selected if the capacity of $S$ is greater (smaller) than that of $R$.

\subsection{Complexity}
The worst--case computational complexity of the sBB algorithm used to solve the offline power allocation scheme for link adaptive relaying is exponential in $K$ \cite{paper47}, whereas the computational complexity of the exhaustive search algorithm is always exponential in $K$. Moreover, the worst case complexity of the sBB algorithm is not likely to occur for all possible realizations of channel SNRs and harvested energies which is evident from the execution time results shown in the simulation results section. Determining the exact and/or average complexity of the sBB algorithm  is more involved and beyond the scope of this paper. The proposed online schemes for link adaptive relaying have linear complexities in $K$.

%\newpage
%\begin{table}[h!]
%%\hrulefill \vspace*{4pt}
%\caption{Online Power Allocation Algorithm For Buffer--Aided Link Adaptive Transmission Protocol} \vspace*{-5mm}
%\label{tab2} \vspace*{-10mm}
%%\hrulefill \vspace*{0pt}
%\end{table}
\begin{algorithm}
\caption{Online Power Allocation Algorithm For Buffer--Aided Link Adaptive Relaying}
%\begin{algorithmic}
\vspace{0.2cm}
1: Initialize the buffer status, $Q_0 = 0$ bits;\\
2: \textbf{for} $k=1$ \textbf{to} $K$ \textbf{do} \\
3: \; \; Calculate $B_{S,k}$ and $B_{R,k}$ using (\ref{eq_new_1_pre}) and (\ref{eq_new_1}), respectively.\\
4: \; \; Calculate $P_{S,k}^*$, $P_{R,k}^*$, and $d_k^*$ using (\ref{eqB3})--(\ref{eqB6_pre}) and (\ref{eqB2A})--(\ref{eqB4A}).\\
5: \; \; \textbf{if} $d_{k}^* = 0$ \textbf{then}\\
6: \; \; \; \; \textbf{if} $P_{S,k}^* > B_{S,k}$ \textbf{then}\\
7: \; \; \; \; \; \; $P_{S,k}^* = B_{S,k}$;\\
8: \; \; \; \; \textbf{end if}\\
9: \; \; \; \; $Q_k = Q_{k-1}+\log_2(1+\gamma_{S,k}P_{S,k}^*)$;\\
10: \, \; \textbf{else}\\
11: \, \; \; \; \textbf{if} $P_{R,k}^* > B_{R,k}$ \textbf{then}\\
12: \, \; \; \; \; \; $P_{R,k}^* = B_{R,k}$;\\
13: \, \; \; \; \textbf{end if}\\
14: \, \; \; \; \textbf{if} $\log_2(1+\gamma_{R,k}P_{R,k}^*) > Q_{k}$ \textbf{then}\\
15: \, \; \; \; \; \; $P_{R,k}^* = \frac{2^{Q_k}-1}{\gamma_{R,k}}$;\\
16: \, \; \; \; \textbf{end if}\\
17: \, \; \; \; $Q_k = Q_{k-1}-\log_2(1+\gamma_{R,k}P_{R,k}^*)$;\\
18: \, \; \textbf{end if}\\
19: \textbf{end for}\\
20: Obtain throughput = $\sum\limits_{k=1}^{K} d_k \log_2(1 + \gamma_{R,k}P_{R,k}^*)$.
%\end{algorithmic}
\end{algorithm}

%$P_{R,k}^* \ne 0$ is valid, if there are at least some packets in the buffer at $R$.

%\input{complexity.tex}

\section{Simulation Results \label{s5}}

In this section, we evaluate the performance of the proposed offline and online power allocation schemes for the conventional and link adaptive relaying protocols. We assume that the (overall) average harvesting rate is $H_{S,E}=H_{R,E}=H_E$, and $H_{S,k}$ and $H_{R,k}$ independently take values from the set $\{0,H_E,2H_E\}$, where all elements of the set are equiprobable. For Figs. \ref{fig1}--\ref{fig10}, \ref{fig12}, and \ref{fig13}, we assume $H_E = 0.5$. We adopt $B_{S,max} = B_{R,max} = B_{max}$, where $B_{max} = 4$ for Figs. \ref{fig1}--\ref{fig4}, \ref{fig11} and $B_{max} = 10$ for Figs.~\ref{fig5}--\ref{fig10}, \ref{fig13}. We assume i.i.d. Rayleigh fading channels with $\bar\gamma_{S}=\bar\gamma_{R} = \bar\gamma$ for Figs.~\ref{fig1}--\ref{fig9} and \ref{fig13} and i.n.d. Rayleigh fading channels for Figs.~\ref{fig10}--\ref{fig12}. We simulate $10^4$ randomly generated realizations of the $S$--$R$ and the $R$--$D$ channels and the harvested energies at $S$ and $R$ to obtain the average throughput.

\subsection{Performance of Different Power Allocation Schemes for Conventional Relaying}
In this subsection, we show the performance of the proposed power allocation schemes for conventional relaying. In particular, the impact of the average channel SNR and the number of time slots on the total number of transmitted bits is studied.

%\newpage
\begin{figure}[h]
\vspace{-6mm}
\centering
\includegraphics[width = 4in]{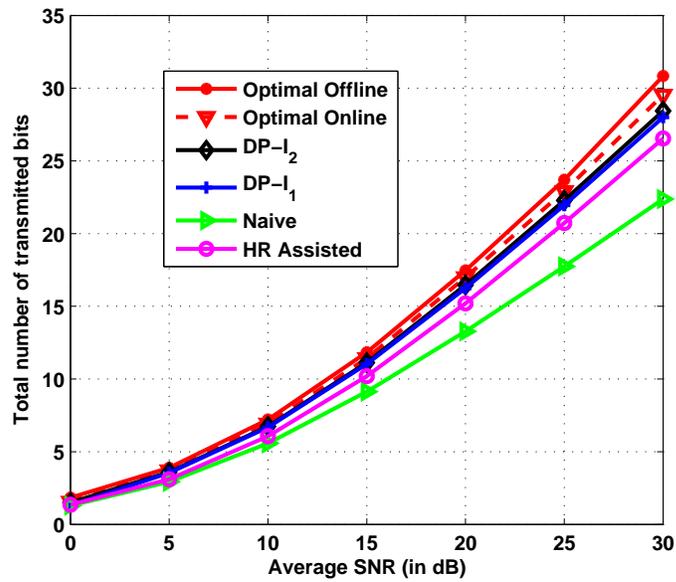} \vspace{-6mm}
 \caption{Conventional relaying: Total number of transmitted bits vs. average channel SNR $\bar\gamma$ for $K=10$. } \label{fig1} \vspace{-6mm}
\end{figure}

%\subsubsection{Total number of transmitted bits vs. $\bar\gamma$}
Fig. \ref{fig1} shows the total number of transmitted bits for the power allocation schemes proposed for conventional relaying vs. the average channel SNR, $\bar\gamma$, for $K=10$. We observe that for all considered schemes, the total throughput increases as $\bar\gamma$ increases. We also notice that the offline scheme performs better than the online power allocation schemes for all $\bar\gamma$. This is due to the fact that in the optimal offline scheme we assume that both causal and non--causal information regarding the CSI and the harvested energy are available whereas the online schemes are based only on causal information regarding the CSI and the harvested energy. Moreover, as expected, the optimal online scheme outperforms all considered suboptimal online schemes and performs close to the optimal offline scheme. The suboptimal online schemes DP--${\rm I}_2$ and DP--${\rm I}_1$ perform close to each other for all $\bar\gamma$. We note that both DP--${\rm I}_2$ and DP--${\rm I}_1$ outperform the HR assisted and the naive schemes.

%As the computational complexity of DP increases rapidly with increasing $K$, we show the performance of the optimal online scheme only for $K=10$, i.e., in Fig. \ref{fig1}. However, for $K=40$ in Fig.~\ref{fig2}, DP--${\rm I}_2$ outperforms DP--${\rm I}_1$ at sufficiently high $\bar\gamma$. The HR assisted scheme performs better than the naive scheme for both $K=10$ and $K=40$ at sufficiently high values of $\bar\gamma$. Moreover, since the transmitted powers are constrained by the average energy harvesting rate, the HR assisted scheme yields better performance for large $K$. Indeed, for $K=40$, the HR assisted scheme outperforms DP--${\rm I}_1$ and achieves a performance similar to that of DP--${\rm I}_2$ at high $\bar\gamma$.

\begin{figure}[h]
\vspace{-6mm}
\centering
\includegraphics[width = 4in]{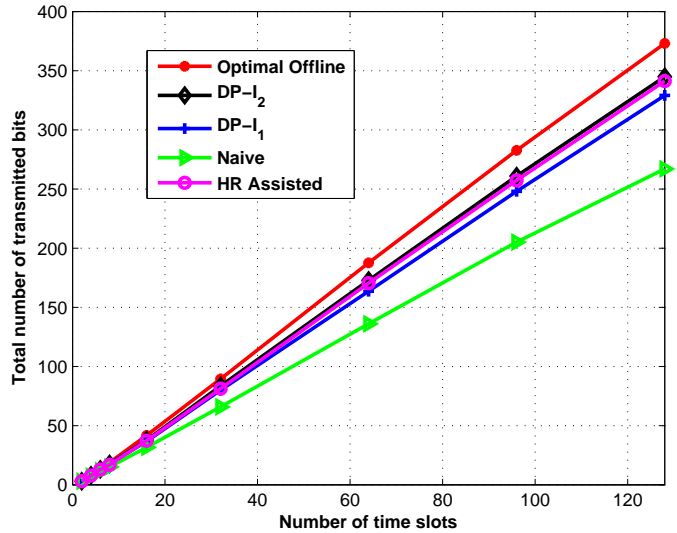} \vspace{-7mm}
 \caption{Conventional relaying: Total number of transmitted bits vs. number of time slots $K$ for $\bar\gamma = 25$ dB. } \label{fig4} \vspace{-6mm}
\end{figure}

%\subsubsection{Total number of transmitted bits vs. $K$}
In Fig. \ref{fig4}, we show the total number of transmitted bits for the power allocation schemes proposed for conventional relaying vs. the number of time slots $K$ for $\bar\gamma=25$ dB. We observe that the optimal offline method achieves the best performance. Among the different suboptimal online schemes, DP--${\rm I}_2$ performs best. The HR assisted scheme provides a similar performance as DP--${\rm I}_2$ for large $K$. This is mainly due to the fact that the HR assisted scheme is based on the average harvesting rate which is a more effective approach for large $K$. Moreover, we observe that the difference between the performances of DP--${\rm I}_2$ and DP--${\rm I}_1$ increases with increasing $K$. This shows that the consideration of the two next time slots instead of only the next time slot for calculation of the optimal transmit powers becomes more important for larger $K$.

\subsection{Performance of Different Power Allocation Schemes for Link Adaptive Relaying}
In this subsection, we show the impact of the average channel SNR and the number of time slots on the total number of transmitted bits for the different power allocation schemes proposed for link adaptive relaying.

\begin{figure}[h]
\vspace{-6mm}
\centering
\includegraphics[width = 4in]{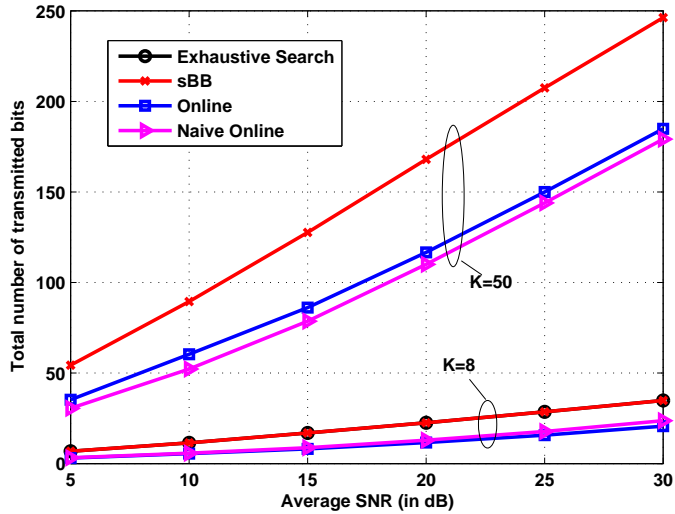} \vspace{-7mm}
 \caption{Link adaptive relaying: Total number of transmitted bits vs. average channel SNR $\bar\gamma$ for $K=8$ and $K=50$. } \label{fig5} \vspace{-6mm}
\end{figure}

Fig. \ref{fig5} shows the total number of transmitted bits for link adaptive relaying vs. the average channel SNR $\bar\gamma$ for $K=8$ and $K=50$. Here, we consider the exhaustive search offline algorithm for $K = 8$, and the online, the naive online, and the sBB offline power allocation schemes for both values of $K$. Recall that the optimal offline exhaustive search scheme is only effective for small $K$ as the complexity increases exponentially with $K$. For $K=8$, we observe that the exhaustive search and the sBB schemes have exactly the same performance for all the considered $\bar\gamma$. This observation confirms that the sBB scheme finds the global optimum solution of non--convex MINLP problems \cite{paper46,paper47}. The performance gap between the offline and the online schemes is small at low $\bar\gamma$ and large at high $\bar\gamma$ for $K=8$. Furthermore, the performance gap increases with $\bar\gamma$ for $K=50$. For $K=8$, the naive online power allocation scheme has a small performance advantage over the online algorithm for high $\bar\gamma$, whereas for $K=50$, the online algorithm shows better performance for all considered $\bar\gamma$.

%In Fig. \ref{fig7} we show the total number of transmitted bits vs. the number of time slots $K$ for $\bar\gamma = 5$ dB and $\bar\gamma = 23$ dB. Due to the increasing complexity of exhaustive search algorithm with increasing $K$, we do not show the performance of this algorithm in this figure. As expected, the sBB scheme achieves better performance than the online scheme for all $K$.

\subsection{Comparison Between Conventional and Link Adaptive Relaying}
In this subsection, we compare the power allocation schemes proposed for conventional and link adaptive relaying. For offline power allocation, we compare the optimal schemes and for online power allocation, we compare the suboptimal schemes for conventional relaying with the online scheme for link adaptive relaying. The optimal DP approach (for conventional relaying) is not included in the comparison because of its high complexity.

\subsubsection{Total number of transmitted bits vs. $K$}

\begin{figure}[h]
\vspace{-6mm}
\centering
\includegraphics[width = 4in]{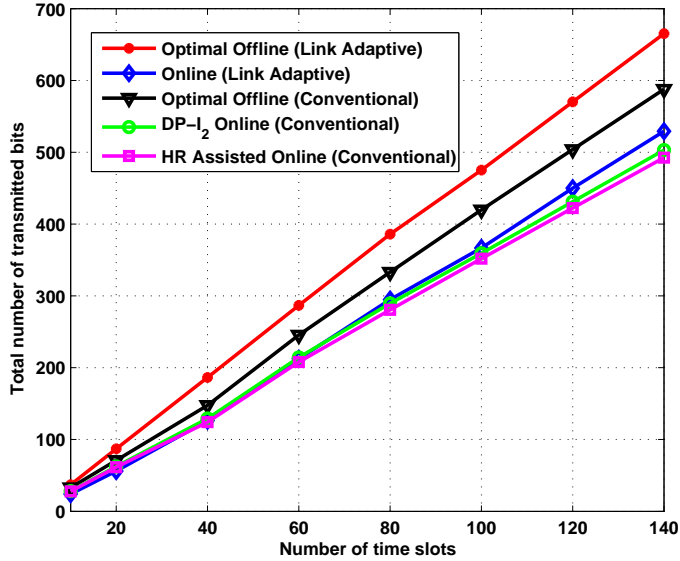} \vspace{-7mm}
 \caption{Comparison of conventional and link adaptive relaying: Total number of transmitted bits vs. number of time slots $K$ for $\bar\gamma_S = \bar\gamma_R = \bar\gamma= 30$ dB. } \label{fig9} \vspace{-6mm}
\end{figure}

Fig.~\ref{fig9} shows the total number of transmitted bits vs. the number of time slots, $K$ for the offline and online power allocation schemes for conventional and link adaptive relaying. We assume symmetric $S$--$R$ and $R$--$D$ channels, i.e., $\bar\gamma_S = \bar\gamma_R = \bar\gamma$. We observe that link adaptive relaying significantly outperforms conventional relaying for offline power allocation. The performance gap increases with increasing $K$. This is mainly due to the fact that, for large $K$, we have more flexibility in selecting $d_k$, i.e., in selecting the $S$--$R$ or $R$--$D$ link for transmission to increase the system throughput. In particular, for the offline case, the link adaptive relaying scheme can transmit 16 and 55 additional bits compared to conventional relaying for $K=20$ and $K=100$, respectively. For small $K$, the online scheme for link adaptive relaying does not show a better performance than the online schemes for conventional relaying. However, for relatively large $K$, e.g. $K=140$, the online scheme for link adaptive relaying outperforms the DP--${\rm I}_2$ and HR assisted online schemes for conventional relaying by 26 and 36 bits, respectively.

\begin{figure}[h]
\vspace{-6mm}
\centering
\includegraphics[width = 4in]{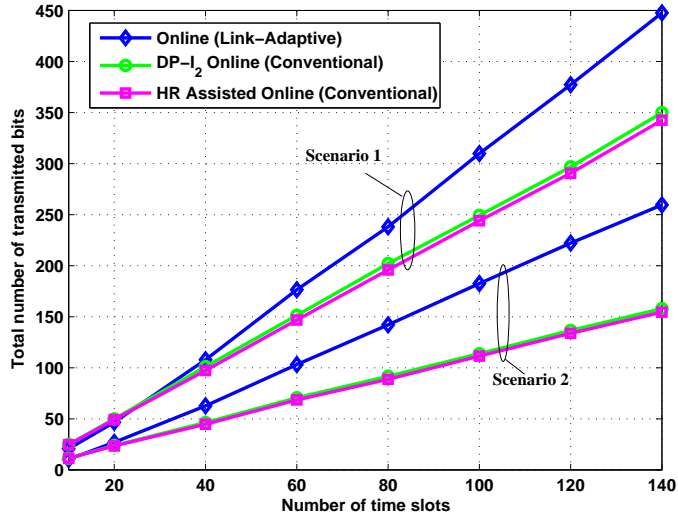} \vspace{-7mm}
 \caption{Comparison of online algorithms for conventional and link adaptive relaying: Total number of transmitted bits vs. number of time slots $K$ for $\bar\gamma_{S}=20$ dB (Scenario 1) and $\bar\gamma_{S} = 10$ dB (Scenario 2) with $\bar\gamma_{R} = 30$ dB. } \label{fig10} \vspace{-6mm}
\end{figure}

In Fig.~\ref{fig10}, we turn our attention to asymmetric links where we consider two scenarios for the average channel SNRs. Scenarios 1 and 2 are valid for $\bar\gamma_{S} = 20 \,\,{\rm dB}$ and $\bar\gamma_{S} = 10 \,\,{\rm dB}$, respectively, where $\bar\gamma_{R} = 30$ dB in both cases. We compare the performance of the online scheme for link adaptive relaying with those of the DP--${\rm I}_2$ and HR assisted schemes for conventional relaying. In contrast to Fig.~\ref{fig9}, we observe that the online scheme for link adaptive relaying outperforms the DP--${\rm I}_2$ (HR assisted) schemes even for small numbers of time slots, e.g. $K=25$. Moreover, Fig.~\ref{fig10} clearly shows that the performance gains of the online scheme for link adaptive relaying over the DP--${\rm I}_2$ (HR assisted) scheme for conventional relaying are significantly larger for asymmetric links compared to symmetric links. The larger gains are caused by the flexibility introduced by the buffer at the relay. In the link adaptive scheme, the stronger link can be used less frequently since relatively large amounts of information can be transferred every time the link is used. Hence, the weaker link can be used more frequently to compensate for its poor link quality. In contrast, in conventional relaying, both links are used for the same amount of time regardless of their respective qualities.

%For instance, for $K=10000$, the online scheme for link adaptive relaying outperforms the DP--${\rm I}_2$ and HR assisted schemes by 4390 and 10040 bits, respectively.

\subsubsection{Total number of transmitted bits vs. $H_E$}

\begin{figure}[h]
\vspace{-6mm}
\centering
\includegraphics[width = 4in]{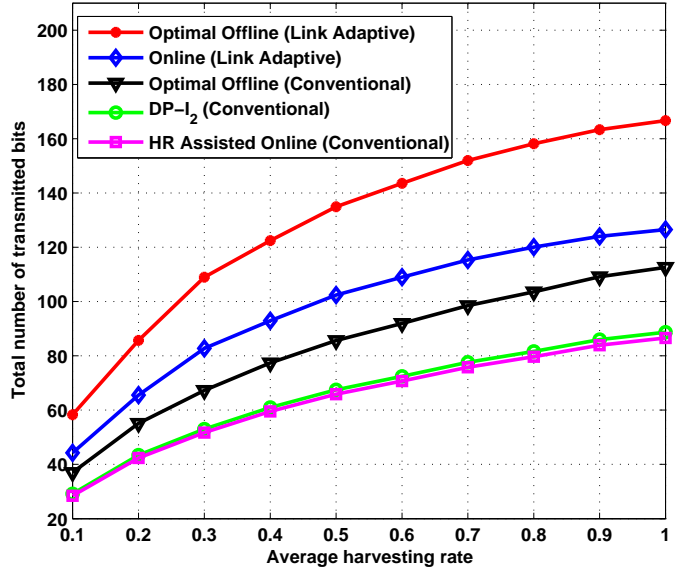} \vspace{-7mm}
 \caption{Comparison of conventional and link adaptive relaying: Total number of transmitted bits vs. $H_E$ for $K=60$, $\bar\gamma_{S}=10$ dB, and $\bar\gamma_{R} = 30$ dB. } \label{fig11} \vspace{-6mm}
\end{figure}

Fig. \ref{fig11} depicts the total number of transmitted bits for conventional and link adaptive relaying vs. the average harvesting rate, $H_E$, for $\bar\gamma_{S}=10$ dB, $\bar\gamma_{R}=30$ dB, and $K=60$. We observe that the throughput increases with increasing $H_E$ for all considered power allocation schemes. We note that the slope of the throughput curves is large for small $H_E$ and decreases with increasing $H_E$. This behavior is partially (apart from the behavior of the $\log(\cdot)$ function) due to the fact that the performance of all schemes is limited by the finite storage capability of the batteries. For large $H_E$, additional energy cannot be stored in the batteries and therefore the extra amount is wasted. We observe that the optimal offline and online schemes for link adaptive relaying outperform the corresponding schemes for conventional relaying for all $H_E$.

\subsubsection{Total number of transmitted bits vs. $B_{max}$}

\begin{figure}[h]
\vspace{-6mm}
\centering
\includegraphics[width = 4in]{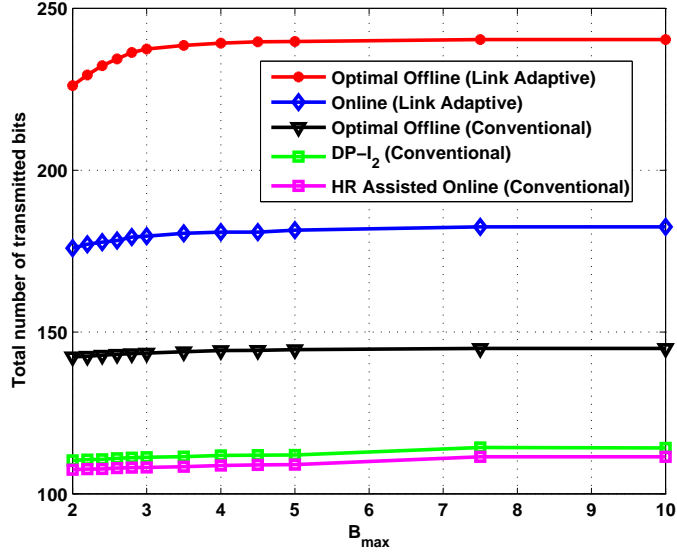} \vspace{-7mm}
 \caption{Comparison of conventional and link adaptive relaying: Total number of transmitted bits vs. $B_{max}$ for $K=100$, $\bar\gamma_{S}=10$ dB, and $\bar\gamma_{R}=30$ dB. } \label{fig12}\vspace{-9mm}
\end{figure}

In Fig. \ref{fig12}, we show the total number of transmitted bits for conventional and link adaptive relaying vs. $B_{max}$ for $\bar\gamma_{S}=10$ dB, $\bar\gamma_{R}=30$ dB, and $K=100$. We observe that for all considered power allocation schemes, the throughput increases with increasing $B_{max}$ and starting at a certain value of $B_{max}$, the throughput remains unchanged. This can be explained by the fact that with $H_E = 0.5$ small values of $B_{max}$ limit the performance since the extra amount of harvested energies cannot be stored in the batteries. However, the constant throughput of all the schemes indicates that, for the given parameters, increasing the storage capacities of the batteries beyond a certain value does not improve the performance of the system. Therefore, Fig.~\ref{fig12} provides an indication for the required storage capacities of the batteries at $S$ and $R$ for different power allocation schemes to achieve a desired performance.

\subsubsection{Execution time vs. $K$}

\begin{figure}[h]
\vspace{-6mm}
\centering
\includegraphics[width = 4in]{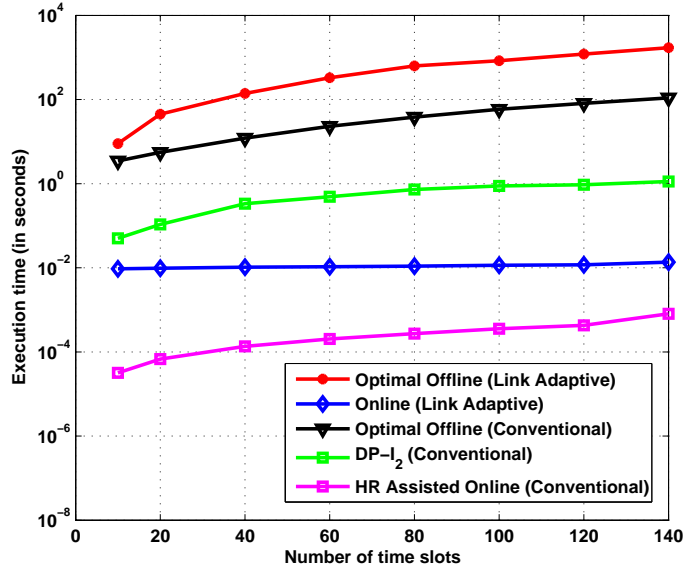} \vspace{-7mm}
 \caption{Comparison of execution times of conventional and link adaptive relaying: Execution time (in seconds) vs. number of time slots $K$ for $\bar\gamma = 30$ dB. } \label{fig13}\vspace{-9mm}
\end{figure}

In Fig.~\ref{fig13}, we show the average execution time (in seconds) vs. the number of time slots, $K$, for all offline and online power allocation schemes for conventional and link adaptive relaying. We ran all the algorithms on the same simulation platform. In particular, all simulations were performed by MATLAB with the Intel(R) Core(TM) i7-2670QM (@2.20GHz 2.20GHz) processor. Therefore, it is justified to compare the complexity of the algorithms based on their execution times. We observe that for both the offline and online schemes for both conventional and link adaptive relaying the execution time increases with $K$. Moreover, the required execution time for the sBB algorithm is higher than that for the offline scheme for the conventional relaying for all $K$. The complexity of forming the look--up tables is not included in this analysis. Fig.~\ref{fig13} provides insightful information about the complexity--performance trade--off of the different proposed algorithms. For example, in case of conventional relaying, the HR assisted scheme is less complex than the DP--${\rm I}_2$ scheme with only a small performance degradation for sufficiently high average channel SNR and large $K$, see Fig.~\ref{fig4}. Therefore, it is preferable to apply the HR assisted scheme compared to the DP--${\rm I}_2$ scheme. Moreover, although the worst case complexity of the sBB algorithm is exponential in $K$, from Fig.~\ref{fig13} we observe that its average complexity is comparable to that of the offline power allocation scheme for conventional relaying. Thus for link adaptive relaying, we can conclude that the sBB algorithm is preferable over the exhaustive search method (which is not shown in Fig.~\ref{fig13} due to its very high execution time).

%As the HR assisted scheme directly depends on the average harvesting rate of $S$ and $R$, its performance is highly influenced by $H_E$. We observe in Fig. \ref{fig5} that the HR assisted scheme shows the best performance among all the suboptimal online schemes for low $H_E$. However, the performance of the HR assisted scheme is somewhat limited for high $H_E$. In fact, the factor, that the harvested energy usage is limited by the maximum storage capacity of the batteries, is more prominent for the HR assisted scheme than for the other schemes.

%%%%%%%%%%%%%%%%%%%%%%%
%
%\begin{figure}[t]
%\centering
%\includegraphics[height=60mm, width = 90mm]{Fig_1_Mar_18.eps}
%\vspace*{-8mm} \caption{BER of 4--PSK vs.~average SNR, $\bar\gamma$, for AF relay selection.}\label{fig1} \vspace*{-7.5mm}
%\end{figure}
%
%%%%%%%%%%%%%%%%%%%%%%%%

\section{Conclusions \label{s6}}

In this paper, we have considered the problem of transmit power allocation for single relay networks, where the source and the relay harvest the energy needed for transmission from the surrounding environment. Two different transmission strategies, namely conventional relaying and link adaptive relaying, have been considered. We have proposed several optimal and suboptimal offline and online power allocation schemes maximizing the system throughput of the considered EH systems over a finite number of time slots. Simulation results showed that the proposed suboptimal online schemes present a good complexity--performance trade--off. Moreover, we showed that, for both offline and online optimization, adopting the link adaptive protocol significantly improves the throughput compared to conventional relaying, especially for asymmetric link qualities.

\renewcommand{\baselinestretch}{0.9}
%\small\normalsize
%\bibliography{litdab}
\bibliographystyle{IEEEtran}
\bibliography{litdab_new}

\begin{thebibliography}{10}
\providecommand{\url}[1]{#1}
\csname url@samestyle\endcsname
\providecommand{\newblock}{\relax}
\providecommand{\bibinfo}[2]{#2}
\providecommand{\BIBentrySTDinterwordspacing}{\spaceskip=0pt\relax}
\providecommand{\BIBentryALTinterwordstretchfactor}{4}
\providecommand{\BIBentryALTinterwordspacing}{\spaceskip=\fontdimen2\font plus
\BIBentryALTinterwordstretchfactor\fontdimen3\font minus
  \fontdimen4\font\relax}
\providecommand{\BIBforeignlanguage}[2]{{%
\expandafter\ifx\csname l@#1\endcsname\relax
\typeout{** WARNING: IEEEtran.bst: No hyphenation pattern has been}%
\typeout{** loaded for the language `#1'. Using the pattern for}%
\typeout{** the default language instead.}%
\else
\language=\csname l@#1\endcsname
\fi
#2}}
\providecommand{\BIBdecl}{\relax}
\BIBdecl

\bibitem{paper_EH_7}
{A.~Kansal, J.~Hsu, S.~Zahedi, and M.~B.~Srivastava}, ``{Power Management in
  Energy Harvesting Sensor Networks},'' \emph{ACM Trans. Embed. Comput. Syst.},
  vol.~6, pp. 1--35, Sep. 2007.

\bibitem{paper_EH_1}
{O.~Ozel, K.~Tutuncuoglu, J.~Yang, S.~Ulukus, and A.~Yener}, ``{Transmission
  with Energy Harvesting Nodes in Fading Wireless Channels: Optimal
  Policies},'' \emph{IEEE J. Select. Areas Commun.}, vol.~29, pp. 1732--1743,
  Sep. 2011.

\bibitem{paper_EH_2}
{C.~K.~Ho and R.~Zhang}, ``{Optimal Energy Allocation for Wireless
  Communications with Energy Harvesting Constraints},'' \emph{[Online]:
  http://arxiv.org/abs/1103.5290}.

\bibitem{paper_EH_3}
{B.~Medepally and N.~B.~Mehta}, ``{Voluntary Energy Harvesting Relays and
  Selection in Cooperative Wireless Networks},'' \emph{IEEE Trans. on Wireless
  Commun.}, vol.~9, pp. 3543--3553, Nov. 2010.

\bibitem{paper_EH_8}
{J.~Yang and S.~Ulukus}, ``{Transmission Completion Time Minimization in an
  Energy Harvesting System},'' \emph{Proceedings of Information Sciences and
  Systems (CISS)}, pp. 1--6, Mar. 2010.

\bibitem{paper_EH_9}
------, ``{Optimal Packet Scheduling in an Energy Harvesting Communication
  System},'' \emph{[Online]: http://arxiv.org/abs/1010.1295}.

\bibitem{paper_EH_11}
{J.~Yang, O.~Ozel, and S.~Ulukus}, ``{Broadcasting with an Energy Harvesting
  Rechargeable Transmitter},'' \emph{{IEEE Trans. Inform. Theory (submitted)
  [Online]: http://www.ece.umd.edu/~ulukus/papers/journal/bc-ener-harv.pdf}}.

\bibitem{paper_EH_6}
{V.~Sharma, U.~Mukherji, V.~Joseph, and S.~Gupta}, ``{Optimal Energy Management
  Policies for Energy Harvesting Sensor Nodes},'' \emph{IEEE Trans. on Wireless
  Commun.}, vol.~9, pp. 1326--1336, Apr. 2010.

\bibitem{paper_EH_14}
{C.~Huang, R.~Zhang, and S.~Cui}, ``{Throughput Maximization for the Gaussian
  Relay Channel with Energy Harvesting Constraints},''
  \emph{[Online]:http://arxiv.org/abs/1109.0724}.

\bibitem{paper51}
------, ``{Outage Minimization in Fading Channels under Energy Harvesting
  Constraints},'' \emph{Proceedings of IEEE International Conference on
  Communications (ICC)}, Jun. 2012.

\bibitem{paper52}
{B.~Gurakan, O.~Ozel, J.~Yang, and S.~Ulukus}, ``{Energy Cooperation in Energy
  Harvesting Wireless Communications},'' \emph{IEEE International Symposium on
  Information Theory (ISIT), Cambridge, MA.}, Jul. 2012.

\bibitem{paper_EH_12}
{A.~Host--Madsen and J.~Zhang}, ``{Capacity Bounds and Power Allocation for
  Wireless Relay Channels},'' \emph{{IEEE Trans. Inform. Theory}}, vol.~51, pp.
  2020--2040, jun 2005.

\bibitem{paper_EH_13}
{J.~N.~Laneman, D.~N.~C.~Tse, and G.~W.~Wornell}, ``{Cooperative Diversity in
  Wireless Networks: Efficient Protocols and Outage Behaviour},'' \emph{IEEE
  Trans. Inform. Theory}, vol.~50, pp. 3062--3080, Dec. 2004.

\bibitem{paper_EH_15}
{A.~Ikhlef, D.~S.~Michalopoulos, and R.~Schober}, ``{Max-Max Relay Selection
  for Relays with Buffers},'' \emph{IEEE Trans. on Wireless Commun.}, vol.~11,
  pp. 1124 -- 1135, Mar. 2012.

\bibitem{paper_EH_16}
{N.~Zlatanov, R.~Schober, and P.~Popovski}, ``{Buffer-Aided Relaying with
  Adaptive Link Selection},'' \emph{Accepted for IEEE J. Select. Areas
  Commun.}, 2012.

\bibitem{paper46}
{E.~M.~B.~Smith and C.~C.~Pantelides}, ``{A Symbolic Reformulation/Spatial
  Branch--and--Bound Algorithm for the Global Optimization of non--convex
  MINLPs},'' \emph{{Elsevier Journal of Computers and Chemical Engineering}},
  vol.~23, pp. 457--478, 1999.

\bibitem{paper47}
{P.~Belotti, J.~Lee, L.~Liberti, F.~Margot, and A.~Waechter}, ``{Branching and
  Bounds Tightening Techniques for Non--Convex MINLP},'' \emph{{Optimization
  Methods and Software}}, vol.~24, pp. 597--634, 2009.

\bibitem{paper49}
``{Couenne, an exact solver for nonconvex MINLPs},''
  \emph{{[Online]:https://projects.coin-or.org/Couenne}}.

\bibitem{paper44}
{S.~Boyd and L.~Vandenberghe}, ``{Convex Optimization},'' \emph{{Cambridge, UK:
  Cambridge University Press}}, 2004.

\bibitem{paper30}
M.~Grant and S.~Boyd, ``{{CVX}: Matlab Software for Disciplined Convex
  Programming, version 1.21},'' \emph{[Online]: http://www.cvxr.com/cvx/}, Apr.
  2011.

\bibitem{paper45}
{D.~P.~Bertsekas}, ``{Dynamic Programming and Optimal Control Vol. 1},''
  \emph{{Belmont, MA: Athens Scientific}}, 1995.

\bibitem{paper_EH_19}
{R.~Corless, G.~Gonnet, D.~Hare, D.~Jeffrey, and D.~Knuth}, ``{On the LambertW
  Function},'' \emph{{Advances in Computational Mathematics}}, vol.~5, pp.
  329--359, 1996.

\end{thebibliography}
\bibliographystyle{unsrt}
%\newpage
%\renewcommand{\baselinestretch}{1.2}
%\small\normalsize

%\input{figures_tables.tex}

\end{document}